\documentclass[twoside,fleqn]{article}
\usepackage{espcrc2,psfig,epsfig}

\newcommand{\be}{\begin{equation}}
\newcommand{\ee}{\end{equation}}

\newcommand{\AmS}{{\protect\the\textfont2
  A\kern-.1667em\lower.5ex\hbox{M}\kern-.125emS}}

\title{Oscillation Solutions to Solar Neutrino Problem $^*$} 
\author{V. Berezinsky\address{INFN, Laboratori Nazionali del Gran Sasso,
             I--67010 Assergi (AQ), Italy\\ and Institute for Nuclear 
Research, Moscow, Russia}
 }       
\begin{document}

\begin{abstract}
The current status of oscillation solutions to the Solar Neutrino Problem is 
reviewed. Four oscillation solutions are discussed in the light of 708d 
Superkamiokande data: MSW, Just-So VO, VO with Energy-Independent Suppression 
(EIS) and Resonant-Spin-Flavor-Precession (RSFP). Only EIS VO is strongly 
disfavoured by the global rates, mostly due to the Homestake data. Vacuum 
oscillations give 
an interesting solution which explains high-energy excess of events observed 
by Superkamiokande and predicts {\em semi-annual} seasonal variation of 
$Be$-neutrino flux. There are indications to these variations in the GALLEX and 
Homestake data. No direct evidence for oscillation is found yet.    

\end{abstract}

\maketitle

\section{\bf Introduction }

With Superkamiokande in operation and SNO in preparation the study of solar 
neutrinos has entered a decisive stage, when neutrino oscillations can 
be directly discovered. At present we know almost certainly that something 
happens to neutrinos either inside the Sun or on the way between the Sun and 
Earth. This knowledge has been mainly provided by solar neurino experiments 
and by helioseismic observations. \\*[1mm] 
1. {\em Helioseismic data} confirm the Standard Solar Models (SSMs) with 
precision  
sufficient for reliable prediction of neutrino fluxes. The seismic data 
(in agreement within a fraction of percent with the SSM ) are valid 
down to radial distance $0.05 R_{\odot}$, where production of $B$- and $Be$ - 
neutrinos has maximum, while the other neutrinos are mostly produced at 
larger distances. At smaller radial distances, where production of 
neutrinos falls down due to decreasing of volume, the seismic data still exist,
though with worse precision.  

Acoustic frequencies comprise the set of seismic data. Nothing in 
physics is measured with greater precision than frequencies. This is why 
seismic measurements give the super-precise data on density and sound speed 
inside the Sun. Within fraction of percent seismically measured density and\\
---------------------\\
{\em $^*$ Plenary talk at 19th Texas Symposium, 1998}
\newpage
\noindent 
sound speed are different from the 
SSM predictions (especially at distance $0.7 R_{\odot}$) and this difference 
is statisticlally significant. It might imply that some physical processes 
are not included in SSM's, and they are of great interest for physics of 
the Sun. But not for solar neutrinos! This statistically significant 
difference, {\em e.g.} in measured and predicted sound speed, produces  
negligible difference in neutrino fluxes, which is out of interest for any 
present (and most probably for any future) solar-neutrino experiment. 

Almost for half century we thought that solar neutrinos with their tremendous 
penetrating power give us the best way to look inside the Sun. We see now 
that seismic observations do it with higher precision, while solar neutrinos 
give the unique information about neutrino properties.\\*[1mm]
2. {\em Nuclear cross-sections} is now the dominat source of uncertainties 
in the calculated solar-neutrino fluxes. The impressive progress 
exists here too.  In the LUNA experiment at Gran Sasso the cross-section 
of one of the most intriguing reaction,$^3He+^3\!\!He \to ^4\!\!He+2p$, was 
measured at energy corresponding to maximum of the Gamow peak in the 
Sun. The famous speculations about solving or ameliorating the SNP due to 
increase of this cross-section at very low energy, have been now honorably 
buried. In the nearest future most of cross-sections relevant to SNP will 
be measured in the LUNA experiment at very low energy. There is also 
considerable progress in calculations of cross-sections and screening 
of nuclear reactions in the solar plasma. A rather exceptional case is 
cross-section of Hep reaction $p+^3\!\!He \to ^4\!\!He+e^++\nu$, in which  
neutrinos with the highest energies are produced. Uncertainties in calculation 
of this cross-section are very large. \\*[1mm] 
3. {\em Solar Neutrino Problem (SNP)} is a deficit of neutrino fluxes ( as 
compared to the SSM prediction) detected in all solar-neutrino experiments 
(Homestake, SAGE, GALLEX, Kamiokande and Superkamiokande). This deficit 
is described by factor $\sim 3$ for Homestake and by factor $\sim 2$ for all 
other experiments.\\*[1mm]
4. {\em Astrophysical solution} to SNP is strongly disfavoured by combination 
of any two solar-neutrino experiments, {\em e.g.} the boron and chlorine 
experiments (Superkamiokande and Homestake) or the gallium and boron 
experiments (GALLEX/SAGE and Superkamiokande). The ratio 
of $Be$ to $B$ neutrino fluxes, extracted from each pair of experiments 
mentioned above, is negative (or too small). This is the essence of failure 
of astrophysical solution. The arbitrary variation of temperature and unknown 
cross-sections do not solve a problem of $Be/B$ ratio.\\*[1mm]
{\em Solar neutrino experiments have a status of disappearance oscillation 
experiment}.\\*[1mm]
But solar-neutrino oscillations are not proved yet. In this paper I will 
discuss the status of different oscillation solutions to SNP.
\section{Status of Astrophysical Solution to SNP}
The global rates of four solar-neutrino experiments 
\cite{Inoue,GALLEX,SAGE,Hom}, as reported up to 1999, are listed in Table 1 
and compared with calculations of Bahcall and Pinsonneault 1998 \cite{BP98}.
\begin{table}[t]  
\caption{
The solar-neutrino data of 1998 compared with the SSM prediction , Bahcall
and Pinnsoneault 1998 \cite{BP98}. The data of Superkamiokande are given 
in units $10^6~cm^{-2}s^{-1}$.}
\vspace{3mm}
\center{\begin{tabular}{||c|c|c|c||}
\hline 
                &  DATA              & SSM [5]             & DATA/SSM \\
\hline
SUPERK\         &                    &                     &             \\
                &$2.42 \pm 0.08$     &5.15          &$0.47\pm 0.02$\\
GALLEX\         &                    &                     &            \\
(SNU)           &$77.5\pm 7.7$       &129          &$0.60\pm0.06$\\
SAGE\           &                    &                     &            \\
(SNU)           &$66.6\pm 8.0$       &129             &$0.52\pm0.06$\\
HOMEST\         &                    &                     &              \\
(SNU)           &$2.56\pm 0.23       $ &7.7               &$0.33\pm0.03$  \\
\hline
\end{tabular}}
\end{table}

The deficit of detected neutrino fluxes seen in the last column of Table 1 
is impossible to explain by astrophysics or/and nuclear physics. This 
conclusion is based on the following.\\*[1mm]
\noindent
(i) Compatibility of the boron (Superkamiokande) and chlorine (Homestake) 
signals or boron and gallium (GALLEX/SAGE) signals results in unphysically 
small ratio $\Phi_{\nu}(Be)/\Phi_{\nu}(B)$ \cite{BaBe}-\cite{HaLa98}. The 
best fit value of this ratio is negative.
The statement above is model-independent.\\*[1mm]
(ii) The arbitrary variation of unknown nuclear cross-sections and the 
central temperature cannot bring $\Phi_{\nu}(Be)/\Phi_{\nu}(B)$ ratio in 
agreement with observations \cite{lasthope}.\\*[1mm]  
(iii) Seismic observations of the density and sound speed radial profiles 
confirm SSM at distances down to $0.05 R_{\odot}$ with accuracy better than 
fraction of percent \cite{Dz}-\cite{Tu-Ch} and at the center with accuracy 
better 
than $4\%$ for sound speed \cite{ferrara1}. \\*[1mm]
(iv) As minimum, SSM is a good approximation to realistic model of the Sun. 
In this case there must be a track, when changing the parameters of SSM and/or 
introducing the new physical phenomena, one arrives from SSM neutrino fluxes 
to the observed ones. Such a track does not exist \cite{HaLa98}.\\*[1mm]
We shall analyze the last item at some details. But two comments are in 
order now.

There are two other, more model-dependent arguments against astrophysical 
solution to SNP.

If one takes three major components of neutrino fluxes  ($pp$, $Be$ and $B$)  
as independent and positive, and the  CNO neutrino flux (which gives much 
smaller contribution) - according to SSM, then arbitrary variation of those 
three fluxes do not give acceptable fit to the observational data at 
99.99$\%$CL \cite{BKS}.

The deficit of $B$ neutrinos seen in the Superkamiokande data (Table 1)
is another 
problem for astrophysical solution. Some time ago many people thought 
that with extreme and correlated uncertainties in $pBe$-cross-section and in 
the central temperature $T_c$  this discrepancy can be eliminated. Now 
the situation looks like follows. In the helioseismically constrained solar 
models (HCSM) \cite{HCSM} the central temperature $T_c= 1.58 \cdot 10^7~K$ 
within maximum  
uncertainty $\Delta T_c/T_c = 1.4 \%$. Taking $T_c$ 1.4$\%$ lower and $S_{17}$ 
40$\%$ lower  we obtain the minimum $B$-neutrino flux 
$3.0 \cdot 10^6~cm^{-2}s^{-1}$, {\em i.e.} 7.4$\sigma$ higher than the 
measured one.
\vspace{-18mm}  
\begin{figure}[htb]
\epsfxsize=9.5truecm
\centerline{\epsffile{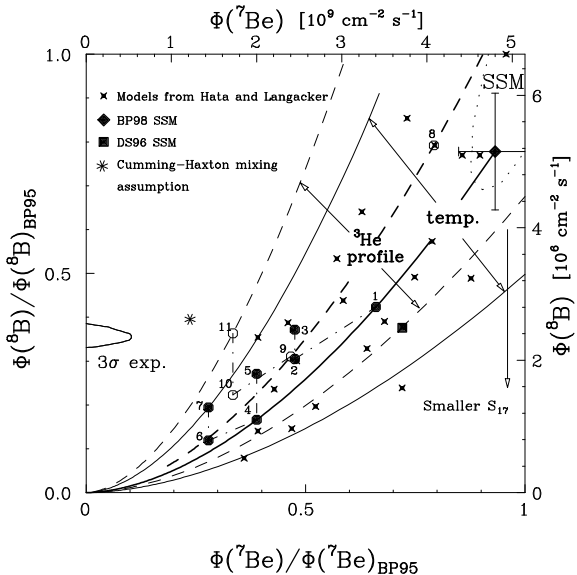}}
\vspace{-40mm}
\caption{\em Neutrino fluxes allowed by arbitrary $^3\!He$ mixing accompanied 
by independent variations of temperature, $S_{34}$ and $S_{17}$. The solid 
lines limit the 
allowed region in case $^3\!He$ radial profile is the same as 
in SSM, and the dashed lines limit the region when both, temperature and 
$^3\!He$ profiles are varying. Some trajectories from the SSM allowed regions 
are 
shown for illustration. The best fit is given by point 11, separated by more 
than $6 \sigma$ from experimentally allowed region (shown as 
''$3 \sigma$ exp.'' in the figure).}
\end{figure}
The most recent attempt \cite{Gough,Schatz,CuHa} 
to reconcile the astrophysical solution with measured 
neutrino fluxes involves an old idea of $^3\!He$-mixing in the solar core. 
 In SSM's $^3\!He$-abundance is very low in the 
$Be,B$-neutrino production zone.  $^3\!He$  is accumulated at much larger 
distance 
$r \sim 0.3R_{\odot}$. It is assumed that due to some process (it could be 
gravity wave induced diffusion \cite{Schatz} or non-linear instability
\cite{Gough}) the ''fresh'' $^3\!He$ is brought into solar core. It could 
happen as the short repeating episodes. Then neutrinoless channel in nuclear 
reactions, $^3\!He + ^3\!He \to ^4\!\!He + 2p$, is enhanced and the 
central temperature $T_c$ decreases too. 

A general analysis of astrophysical solution, which includes arbitrary 
$^3\!He$-mixing has been recently performed in ref.(\cite{BFL99}). 
The $^3\!He$-mixing was assumed not to be accompanied by mixing of other 
elements.
Additionally all other relevant parameters in the neutrino production zone 
were varying within wide range: $T_c$ - within $\pm 5\%$, $S_{17}$ - within
$\pm 40\%$ and $S_{34}$ - in the range $(-20\% +40\%)$. The temperature and 
$^3\!He$ radial profiles were also varying. The results are presented in 
Fig.1 as allowed regions between two limiting curves, thin solid ones 
(''temp.'')
and two broken ones (''$^3\!He$ profile''). The best fit is at  
least $6 \sigma$ away from observationally allowed region. It can be 
interpreted as well that there is no allowed track from the SSM's region 
(''SSM'') to the observationally allowed region (see Fig.1). A trajectory with 
variation of temperature $T_c$ is shown by thick solid line (''temp.'')

\section{Oscillation Solutions}
Due to oscillations, electron neutrino emitted from the Sun can be found at 
the Earth as neutrino with another flavor: muon, tau or 
sterile neutrino. These neutrinos either do not give a signal in the detector 
({\em e.g.} muon neutrinos in gallium or chlorine detectors) or interact 
weaker due to NC ({\em e.g.} muon- or tau-neutrinos in Superkamiokande). 
I will not discuss in this review sterile neutrinos. Atmospheric neutrino 
oscillations imply that $\nu_{\mu}$ and $\nu_{\tau}$ neutrinos are maximally 
mixed. In this case solar $\nu_e$ neutrino oscillates with equal probability 
to each of those neutrinos. 

The probability 
to find emitted electron neutrino in the same flavor state in the detector 
$P_{\nu_e \to \nu_e}$ is called survival probability or (less precisely)
suppression factor. 
In general case survival probability depends on energy, 
$P_{\nu_e \to \nu_e}(E)$, {\em i.e.} solar neutrinos are suppressed in 
energy-dependent way and actually this property allows to 
solve SNP with help of oscillations.  
 
Four oscillation solutions are currently discussed in the literature 
(see Table 2.)
\vspace{-3mm}
\begin{table}[htb]  
\caption{
Oscillation Solutions to SNP.}
\center{\begin{tabular}{|c|c|c|}
\hline 
                &   $\sin^2 2\theta$             &  $\Delta m^2$ ($eV^2$)\\   
                &      best fit                  &           best fit    \\ 
\hline               
MSW $\;\;\;\;$\          &                                    &           \\   
$\;\;\;$ SMA    & $ 6.0\cdot 10^{-3}$             &  $5.4\cdot 10^{-6}$\\      
$\;\;\;$ LMA    & $ 0.76$                         &  $1.8\cdot 10^{-5}$\\
$\;\;\;$ LOW    & $ 0.96$                         &  $7.9\cdot 10^{-8}$\\
VO $\;\;\;\;\;\;\;$\              &                                    &            \\
$\;\;\;$Just-so & $0.75$                          &  $8.0\cdot 10^{-11}$\\
$\;\;$EIS     & $\sim 1$                        &  $ 10^{-9} - 10^{-3}$\\
RSFP $\;\;\;$            &                                    &             \\
                & small                           &  $ 10^{-8} - 10^{-7}$\\    
\hline   
\end{tabular}}
\end{table}
\par 
\vspace*{-0.5cm}
1. {\em MSW solution} \cite{MSW}.\\
MSW effect in the Sun is a resonance conversion of $\nu_e$ into $\nu_{\mu}$ or 
$\nu_{\tau}$. For neutrino energies at interest 
it occurs in the narrow layer, $\Delta R \sim 0.01 R_{\odot}$, at the distance 
$R \sim 0.1 R_{\odot}$ from the center of the Sun. There are three MSW 
solutions to SNP , which explain the global rates in all four solar-neutrino 
experiments: Small Mixing Angle (SMA) MSW, Large Mixing Angle (LMA) MSW and 
LOW solution with low probability (it appears only at $99\%CL$). The best fits 
of these solutions to the rates are reported in Table 2.  

2. {\em Vacuum Oscillations (VO)}\\
The concept of vacuum oscillations was first put forward by B.Pontecorvo
\cite{Pont} (for a review see \cite {BiPe}). The survival probability for 
$\nu_e$ neutrino with energy $E$ at distance $r$ is given by 
\be
P_{\nu_e\to\nu_e}= 1 - \sin^2 2\theta \sin^2 \left( \frac{\Delta m^2}
{4E} r \right),
\label{eq:srv}
\ee
where $l_v= 4\pi E/\Delta m^2$ is the vacuum oscillation length. At 
$\Delta m^2 = 8\cdot 10^{-11}~eV^2$  (the best fit) the oscillation length of 
neutrino with energy $E_{\nu} \sim 3~MeV$ is 
$\l_{\nu} \sim 1\cdot 10^{13}~cm$, {\em i.e.} of order of distance between the 
Sun and Earth. That is why this VO solution is called {\em just-so}. Since 
observational data need large suppression of neutrino flux, by factor 
$\sim 2$, $\sin^2 2\theta \sim 1$ is needed: see  Eq.(\ref{eq:srv}). Thus 
just-so VO solution must be large mixing angle solution. The best fit values 
are given in Table 2.

3. {\em EIS VO solution}\\
VO with Energy Independent Suppression (EIS) occurs when 
$\Delta m^2 \gg 10^{-10}~eV^2$. In this case the oscillation length is 
much smaller than distance between the Sun and Earth. Oscillatory function 
in Eq.(\ref{eq:srv}) is averaged to factor 1/2 and hence suppression factor is 
energy independent and equals to  
$1-\frac{1}{2}\sin^2 2\theta$. Since this suppression should be of order 0.5, 
$\;\; \sin^2 2\theta \sim 1$ is needed. On the other hand one must assume 
$\Delta m^2 \ll 10^{-3}~eV^2$ because of non-observation  of 
$\nu_e$ oscillation in the atmospheric neutrinos.  

The energy independent suppression 
is excluded by observed rates at $99.8\% CL$ \cite{BKS}. However, the 
Homestake data give the dominant contribution to this conclusion. If these 
data are  arbitrarily  excluded from analysis, EIS VO survives. I will not 
give more
attention to discussion of EIS VO solution. Further details a reader can 
find in references \cite{Perk,FoVo,Comf,Gold}.   

4. {\em RSFP solution}\\
The Resonant Spin-Flavor Precession (RSFP) describes two physical effects
working simultaneously: the spin-flavor precession, when neutrino spin 
(coupled to magnetic moment)
precesses around magnetic field, changing simultaneously neutrino flavor, and 
the resonant, density-dependent effect, which produces difference in potential 
energy of 
neutrinos with different flavors (similar to the MSW effect). This complex  
transition occurs in the external magnetic field due to presence of 
non-diagonal (transition) neutrino magnetic moments. The RSFP was first 
recognized 
in ref.'s \cite{Akh,Lim}. For excellent review see \cite{Akh-rev}. 

This theory had a predecessor. The precession of neutrino magnetic 
moment around magnetic field converts left-chiral electron neutrino $\nu_{eL}$
into sterile right component $\nu_{eR}$, suppressing thus 
$\nu_e$-flux\cite{VVO}. However, the suppression effect in this case is 
energy-independent and thus contradicts to the observed solar-neutrino rates.
In ref's \cite{VVO,BaFi} the matter effect was included and in \cite{SchVa}
spin-flavor precession was discovered.
The observed rates in 
solar-neutrino experiments can be explained only by the RSFP, because 
only this type of precession  give the energy-dependent suppression factor. 
 Majorana neutrino can have only transition magnetic moment. RSFP induces
the transition $\nu_{eL}$ to $\bar{\nu}_{\mu R}$, i.e. electron neutrino 
to muon antineutrino, which can scatter off the electron due to NC. The 
survival 
probability is similar to that of SMA MSW (see Fig.3 from \cite{Nun}). 
Neutrino mixing is not needed 
directly for RSFP effect, but it is needed indirectly to provide the 
transition magnetic moment of the Majorana neutrino. To be a solution to SNP, 
RSFP needs a transition magnetic moment $\mu \sim 10^{-11} \mu_B$, magnetic 
field in the resonance layer of the Sun, $B \sim 20 - 100~kG$ and 
$\Delta m^2$ in the range $10^{-8} - 10^{-7}~eV^2$ (see Section 6).

\section{Signatures of Oscillation Solutions}

A common signature of most neutrino oscillation solutions  is distortion of 
$B$-neutrino spectrum. The survival probabilities for SMA MSW, LMA MSW 
and just-so VO are shown in Fig.2. The survival probabilities for RSFP are 
similar to SMA MSW  and shown in Fig.3. One can see there that LMA MSW 
(and LOW too) predicts small distortion of $B$-neutrino spectrum spectrum in 
the region of observation $5- 15~MeV$. For EIS VO the distortion is absent. 
The strongest spectrum distortion one can expect for SMA MSW and just-so VO. 
However, spectrum of recoil electrons are distorted weaker than that of 
neutrinos, because of 
cross-section and averaging over energy bins in observations ({\em e.g.} 
see Fig.8).  {\em The absence of distortion of neutrino or recoil-electron 
spectra is not a general argument against neutrino oscillations}.

{\em Anomalous NC/CC ratio} is another common signature of neutrino 
oscillations which can be observed in SNO. The NC events will be seen 
there by detection of neutrons produced in $\nu+D \to p+n+\nu$ reaction. 
Oscillation $\nu_e \to \nu_{\mu} (\nu_{\tau})$ does not change NC interaction
but changes CC 
\newpage
\begin{figure}[t]
\begin{center}
\psfig{bbllx= 7pt, bblly=50pt, bburx=550pt, bbury=720pt,
file=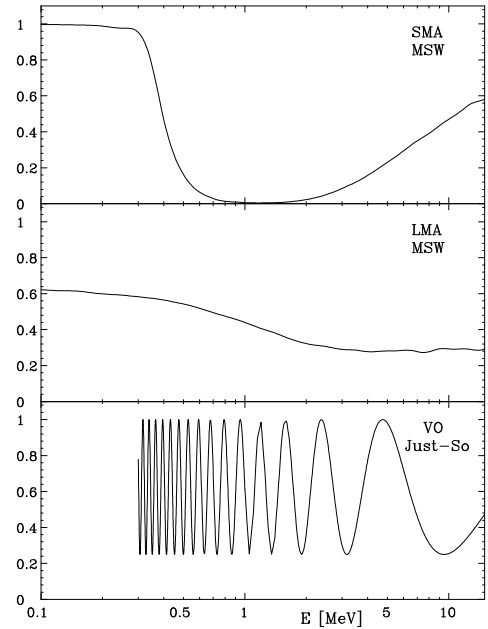, height=8.5cm , clip=}
\end{center}
\vspace{-15mm}
\caption{\em Electron neutrino survival probabilities.}
\end{figure}
\vspace*{-10mm}
\samepage
\vspace*{-10mm}
\begin{figure}[h]
\epsfxsize=7.6truecm
\centerline{\epsffile{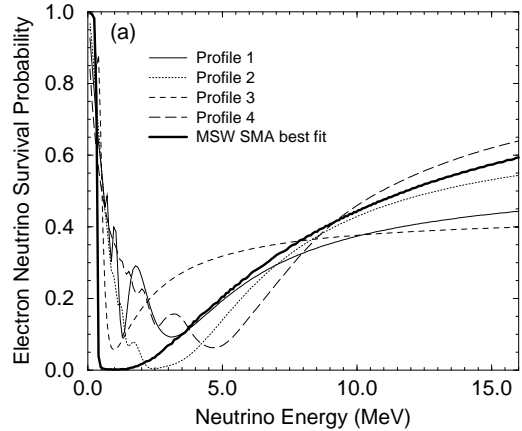}}
\vspace{-10mm}
\caption{\em RSFP survival probabilities from \cite{Nun} compared with SMA MSW 
(best fit) survival probability, shown by thick solid line. The four other 
curves correspond to different radial profiles of magnetic field.}
\vspace*{-60mm}
\end{figure}  
\vspace*{-10mm}   
\newpage
\noindent
interaction and thus the ratio of NC/CC rate. In case of 
oscillation to sterile neutrino the NC/CC ratio is not changed. Therefore,
{\em the normal ratio NC/CC is not a general argument against neutrino 
oscillations}. 

{\em MSW solutions} have very distinct signatures. 
They are day/night effect, zenith angle dependence of solar-neutrino flux 
and difference in day/night neutrino spectra. All these effects are caused 
by MSW matter effect in the Earth. Another related effect is seasonal
variation of neutrino flux caused by longer nights in winters. This
effect is smaller than integrated day/night effect. For recent
calculations see \cite{Valle}  

The signature of {\em just-so VO} is anomalous seasonal variation of neutrino 
flux \cite{Pom}. Due to ellipticity of the Earth's orbit, the distance between 
the Sun and Earth changes with time, causing 
$7\%$ variation of the flux due to $r^{-2}$ effect
(''geometrical'' seasonal variation). Due to just-so 
VO the flux of $\nu_e$ neutrinos changes additionally due to dependence of 
survival probability (\ref{eq:srv}) on distance. As follows from 
Eq.(\ref{eq:srv}) neutrinos with different $E$ have different phases and it 
weakens the observed effect, when averaged over interval $\Delta E$. 
In case of monochromatic $Be$ neutrinos VO seasonal time variations 
are strongest 
For the detailed calculations of anomalous seasonal variations 
see \cite{KrPe} and for the recent calculations 
\cite{GlKe,MiSm,GeRo,Sm,Fo,MaPe}.
  
For {\em EIS VO solution} the anomalous seasonal variations are absent, 
because the oscillation length is too small. It results in a signature, which 
can be observed by BOREXINO: $Be$-neutrino flux is suppressed by a factor 
$\sim 2$, but does not show anomalous seasonal variations. 
\samepage

{\em RSFP} has two signatures.
As a result of RSFP electron neutrinos oscillate inside the Sun into 
muon/tau antineutrinos. Due to vacuum oscillations on the way to the Earth
these neutrinos oscillate to electron antineutrinos. The latter oscillations 
are suppressed by mixing angle, which is small in case of RSFP. However, even 
small fluxes of electron antineutrino can be reliably detected ({\em e.g.} by 
KamLand \cite{ASuz}). 
\samepage
The  second signature is prediction of 11-year periodicity for
$Be$-neutrino flux \cite{Akh-rev}.\\
\newpage
\noindent 
 RSFP occurs in the resonant layers, 
which are located at different distances for neutrinos of different 
energies: for $pp$ and $B$ neutrinos the resonant 
layers are located near the solar center and 
at the periphery, respectively, where magnetic field is weak and RSFP too. 
The resonant layer for $Be$ neutrinos is located at  intermediate distance, 
where magnetic field is large and RSFP is strongest. 11-year variations of 
neutrino flux is caused by periodic variation of toroidal magnetic field 
at the bottom of convective zone. The magnetic activity of the sun  exhibits 
quasi-periodic time variations with the mean period $11~yr$. 
This periodicity 
is thought to be originated due to toroidal field, generated in so-called
{\em overshoot layer} by dynamo mechanism and located near the bottom of 
convective zone. Theoretically, magnetic field there can reach $100~kG$.
This field rises through convective zone to the surface of the sun.
The $11~yr$ periodicity should be observed most effectively by neutrino 
detectors sensitive to $Be$-neutrinos: Homestake, GALLEX and BOREXINO.
In particular (see Fig.3) when toroidal magnetic field disappears 
(due to change of magnetic polarity) survival probability increases from 
$\sim 0.1$ to $\sim 1$. Since $B$-neutrino flux is also suppressed 
by factor $\sim 0.4 - 0.6$, $11~yr$ variations should be seen in the combined  
Kamiokande and Superkamiokande data.
  
\section{708-day  Superkamiokande Data}
After 708 days of solar-neutrino observations Superkamiokande has not found 
direct evidences for neutrino oscillations. There are only some indications 
to the distortion of the spectrum of recoil electrons, which will be discussed 
in this Section. 
\samepage

{\em The spectrum of the recoil electrons} (708 d) is shown in 
Fig.4 \cite{Inoue,Tots} as the ratio to (undistorted) spectrum calculated 
in BP98 SSM \cite{BP98}. The spectrum is suppressed by overall factor 
0.47, but there is no distortion of the  spectrum, except the high energy 
excess at $E_e \geq 13~MeV$.
In principle, this excess can be a result of low statistics 
or small systematic errors at the end of the boron neutrino spectrum. 
For example, due to 
very steep end of the electron spectrum, even small systematic   
\newpage
\noindent
error in electron energy (e.g. due 
to calibration) could enhance the number of events in the highest energy bins.

Another possible explanation of this excess~\cite{BaKr} is that the 
Hep neutrino flux might be significantly larger (about a factor 10--20) 
than the
SSM prediction. The Hep flux depends on solar properties, such as 
$^3$He abundance and  the temperature, and on $S_{13}$, the zero-energy 
astrophysical $S$-factor of the 
\mbox{$p+{}^3He \rightarrow {}^4He + e^+ + \nu$}
reaction. Both SSM based~\cite{BaKr} and model-independent~\cite{BDFR}
approaches give a robust prediction for the ratio $\Phi_{\nu}(Hep)/S_{13}$.
Therefore, this scenario implies a cross-section larger by
a factor 10--20  than the present calculations.
Such a large correction to the calculation
does not seem likely, though is not excluded. 
The signature of Hep neutrinos, the presence of
electrons above the maximum boron neutrino energy, can be tested by
the SNO experiment.

The observed excess is difficult to explain by neutrino oscillations. The 
oscillation parameters $\sin^2 2\theta$ and $\Delta m^2$, which correspond to 
the allowed global rates in four neutrino experiments, result in the recoil  
electron spectra in bad agreement with the excess. The spectra for the best 
fit MSW solutions (LMA shown by short-dash lines and SMA -- by long-dash 
lines) are displayed in Fig.4 (calculations by K.Inoue). 

{\em Night/Day excess} after 201.6, 504 and 708 days of observations,
is found to be small, consistent with zero within 
$1.7 \sigma$:  
\vspace{-3mm}
$$  
 \frac{N-D}{N+D}\times 100 =- 0.4 \pm 3.1 \pm 1.7 \;\;\;\;\; (201.6 d)
$$
\vspace{-3mm}
$$
 \frac{N-D}{N+D}\times 100 =+ 2.3 \pm 2.0 \pm 1.4 \;\;\;\;\; (504 d)
$$
\vspace{-3mm}
$$
 \frac{N-D}{N+D}\times 100 =+ 2.9 \pm 1.7  \pm 0.39\;\;\;\;\; (708 d)
$$
The observed (708 d) excess has the right sign, but is statistically 
insignificant ($1.7\sigma$). Note that systematic error is much smaller 
than observed effect. Statistics, if increased by factor 5, can make the 
effect statistically significant. For clear discussion see \cite{Lisi}. 
{\em Absence of day/night effect does not rule out MSW solution.} 

The zenith-angle dependence is not seen in Superkamiokande data. 
\newpage

The Superkamiokande data for day/night effect, zenith-angle dependence 
and recoil-electron energy spectrum (especially absence of distortion in most 
part of the spectrum) have the great restriction power for the  oscillation 
solutions.     
\vspace*{-5mm}
\begin{figure}[h]
\begin{center}
\psfig{bbllx= 55pt, bblly=175pt, bburx=540pt, bbury=655pt,
file=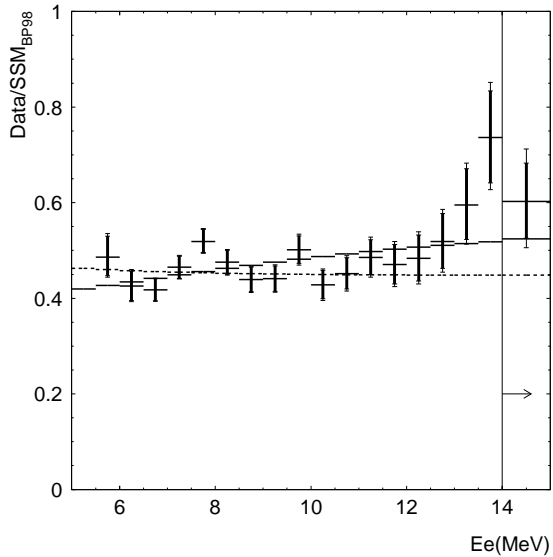, height=7.5cm , clip=}
\end{center}
\vspace{-13mm}
\caption{\em Energy spectrum of recoil electrons from 708d of Superkamiokande 
data \cite{Inoue,Tots}. Plotted is ratio of the observed spectrum to one 
predicted by BP SSM \cite{BP98}. The long- and short- dash lines show SMA 
MSW and LMA MSW spectra, respectively, for best-fit rates (calculations 
by K.Inoue).}
\end{figure}
\vspace*{-10mm}
\section{Status of Oscillation Solutions}
I will discuss here the status of MSW, VO and RSFP solutions in the light of 
708d Superkamiokande data.\\*[1mm]
{\bf 6.1 MSW solutions}\\*[1mm]
The regions in oscillation parameter space allowed by global rates do not 
explain the high energy excess in the recoil-electron spectrum \cite{Suz}
(see Fig.4).
MSW solutions need an alternative explanation of this excess, {\em e.g.} by 
Hep neutrinos (see Section 5). The status of MSW solutions is determined by 
{\em combined} restrictions 
due to global rates, day/night effect, zenith-angle flux dependence and 
the energy spectrum (under
\newpage
\noindent
assumption of arbitrary $S_{13}$). The result 
of such analysis is expressed in goodness of the total fit $\chi^2$. 
\vspace{-13mm}  
\begin{figure}[h]
\epsfxsize=11.5truecm
\centerline{\epsffile{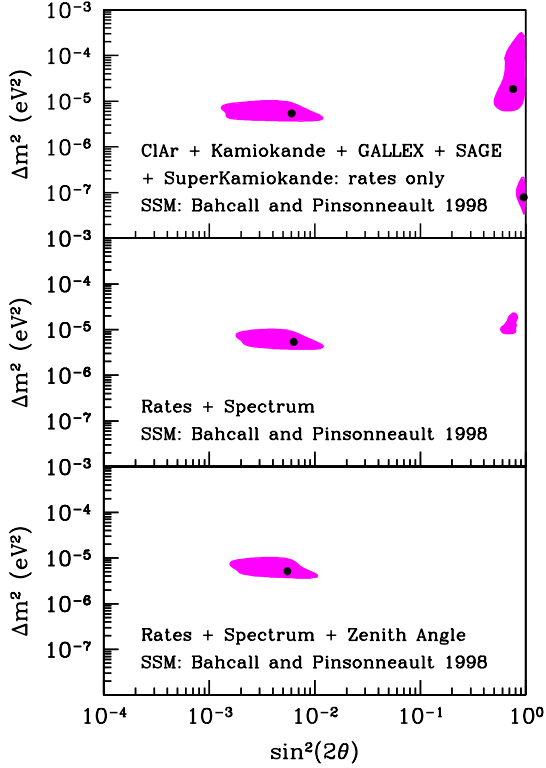}}
\vspace{-15mm}
\caption{\em Status of MSW oscillation solutions after 504 days of 
Superkamiokande data \cite{BKS}.} 
\end{figure}
\vspace{-5mm}

The data of Superkamiokande for 504 days disfavoured LMA MSW solution 
\cite{BKS}. In Fig.5 the upper panel shows the regions allowed (at $99\%$ CL)
by the global rates. The middle panel includes additionally restrictions 
due to spectrum and the low one includes three restrictions (rates, 
spectrum and zenith-angle dependence). All regions are shown at $99\%$ CL.
One can see how LMA and LOW solutions disappear. 

In Fig.6 the 708d data of Superkamiokande are shown as allowed 
by rates (upper panel), the regions excluded by day/night effect 
(middle panel) and excluded by spectrum (low panel). Note, that in the middle 
panel the regions above and below the central one are allowed at $68\%$ CL. 
\newpage
\begin{figure}[ht]
\begin{center}
\includegraphics[bbllx=30pt,bblly=154pt,bburx=535pt,bbury=660pt,
height=50mm,width=65mm]{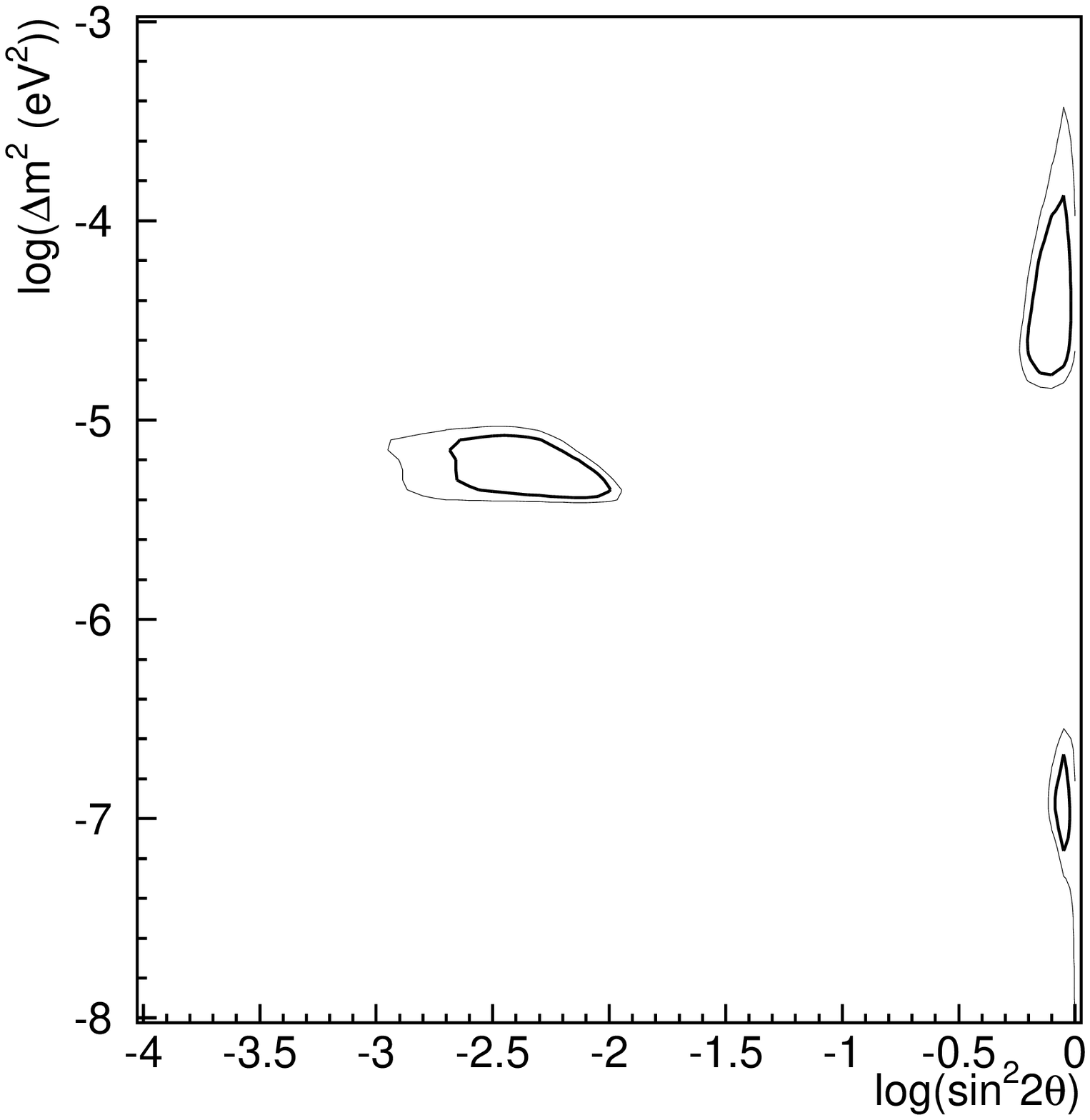}
\end{center}
\vspace{-2mm}
\begin{center}
\includegraphics[bbllx=30pt,bblly=154pt,bburx=535pt,bbury=660pt,
height=50mm,width=65mm]{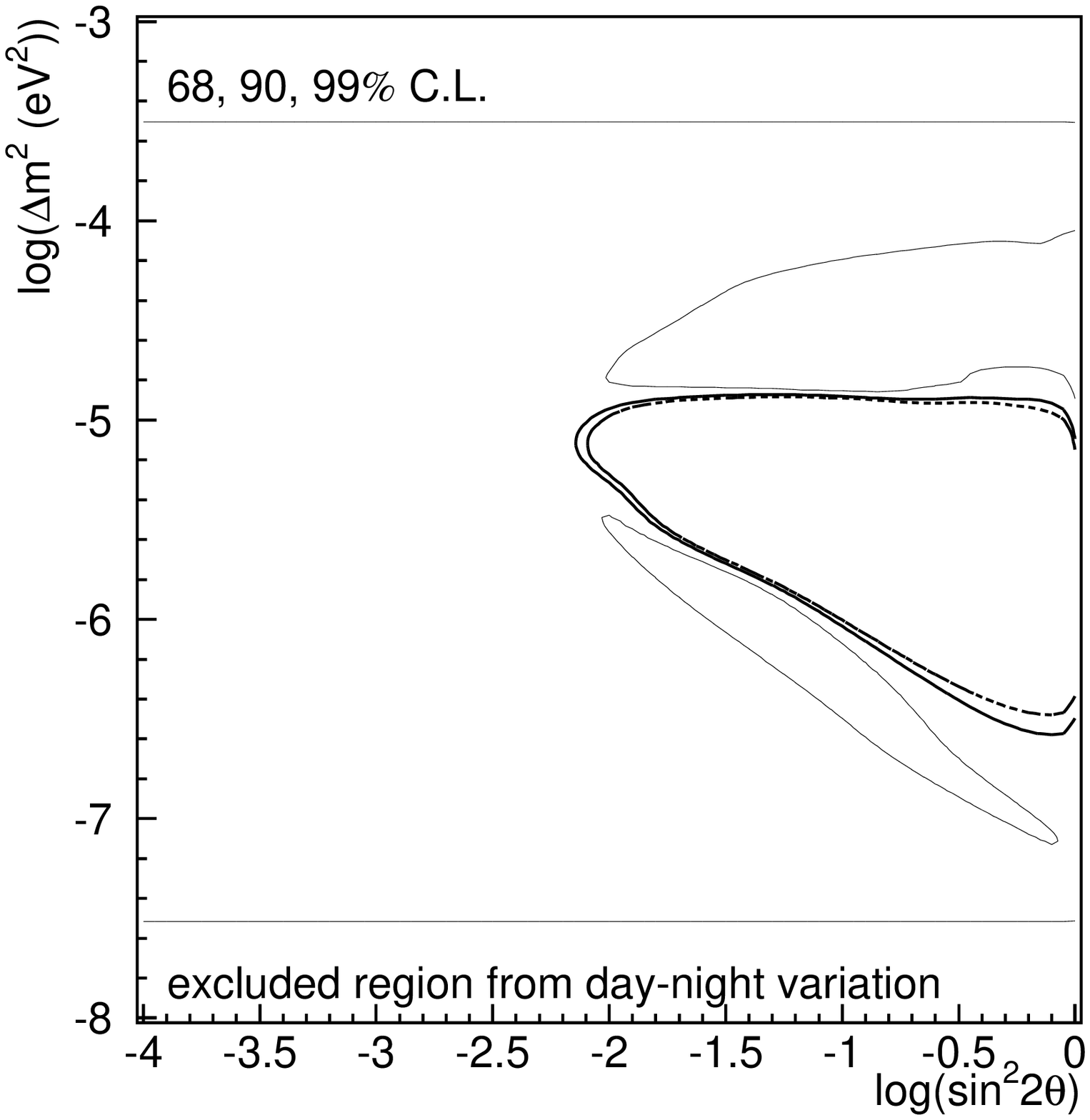}
\end{center}
\vspace{-2mm}
\begin{center}
\includegraphics[bbllx=30pt,bblly=154pt,bburx=535pt,bbury=660pt,
height=50mm,width=65mm]{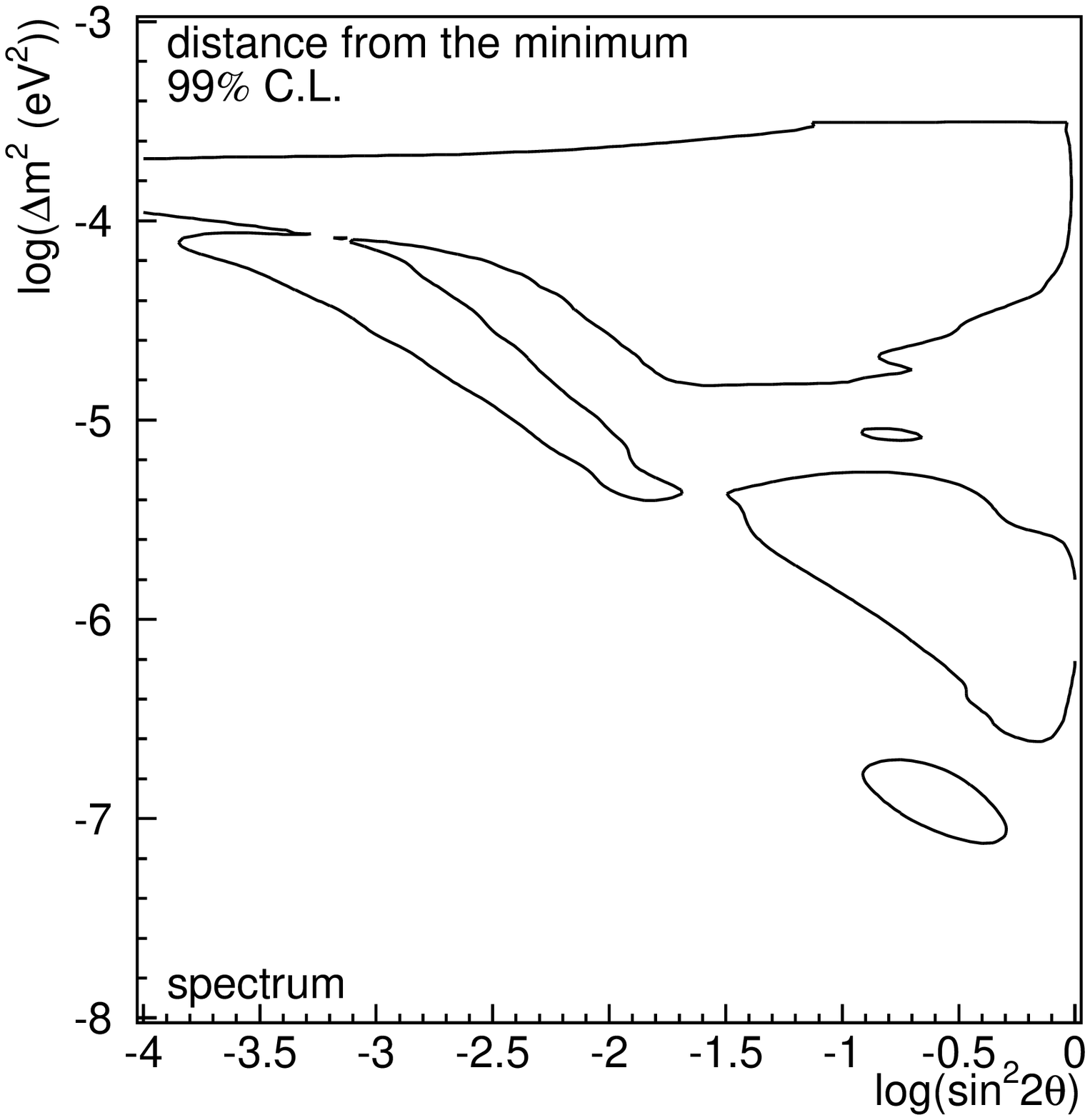}
\end{center}
\vspace{-12mm}
\caption{\em Status of MSW solutions after 708 days of Superkamiokande data 
(courtesy of Y.Totsuka)}
\vspace{-20mm}
\end{figure}
\vspace{-40mm}
\newpage
\noindent
The visual inspection shows that all three MSW solutions 
are allowed.    

However, the  quantitative analysis of 708d data presented by 
K.Inoue \cite{Inoue} shows that the SMA MSW solution is not acceptable at 
$90\%$ CL if day/night effect and 
spectrum (with free Hep flux) are included in the analysis simultaneously.
LMA MSW solution is more favourable.

In conclusion, exclusion of any MSW solution looks unstable and 
statistics-dependent. I mean that conclusions are changing too drastically
with accumulation of data and with method of analysis (inclusion 
or not day and night spectra, inclusion Hep flux as a free parameter 
{\em etc}). Inclusion of too many data together may be misleading if 
the data are partially inconsistent. Finally, I would like to remind a reader 
that 
the data of Superkamiokande are still preliminary, and conclude that it is 
premature to speak of exclusion of any MSW solutions.\\*[1mm]
{\bf 6.2 VO Solutions}\\*[1mm] 
If high energy excess in the spectrum is due to Hep neutrinos, just-so VO
solution fits the rates and the spectrum. In case the excess is due to 
oscillations, the regions in oscillation parameter space 
allowed by the rates (Fig.7, upper panel) are excluded by energy spectrum 
(low panel).
The spectrum is well fitted by vacuum oscillations with  
$\Delta m^2=4.2\cdot 10^{-10}~eV^2$ and $sin^2 2\theta=0.93$ \cite{Suz}, 
but this point is located outside the regions allowed by the rates (Fig.7), 
{\em i.e.} it does not represent the SNP solution. The status of this 
point has been further analysed in \cite{BFL98}.   

To explain both the excess and the rates it was assumed that 
boron neutrino flux is 15--20\% smaller than the SSM 
prediction, and  that the chlorine signal is about 30\%
larger than the Homestake observation.
This assumed $3.4\sigma$ increase of the chlorine signal could have a 
combined statistical and systematic origin. In practice, the SSM boron 
neutrino flux and the Homestake signal were rescaled with help of parameters 
$f_B$ and $f_{Cl}$, as $\Phi_B= f_B \Phi_B^{SSM}$ and 
$R_{Cl} = 2.56 f_{Cl}$~SNU, where 2.56~SNU is the Homestake signal.
\newpage
For each pair $f_B$ and $f_{Cl}$ there were found the parameters
($\Delta m^2, \sin^22\theta$), that explain
the observed rates, and $B$-neutrino spectrum was calculated for these 
parameter values. 
\samepage
\begin{figure}[h]
\begin{center}
\includegraphics[bbllx= 15pt, bblly=150pt, bburx=540pt, bbury=660pt,
height=60mm,width=75mm]{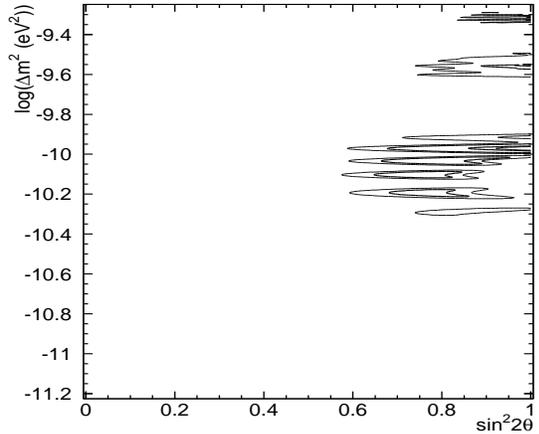}
\end{center}
\begin{center} 
\includegraphics[bbllx= 15pt, bblly=150pt, bburx=530pt, bbury=660pt,
height=60mm,width=75mm]{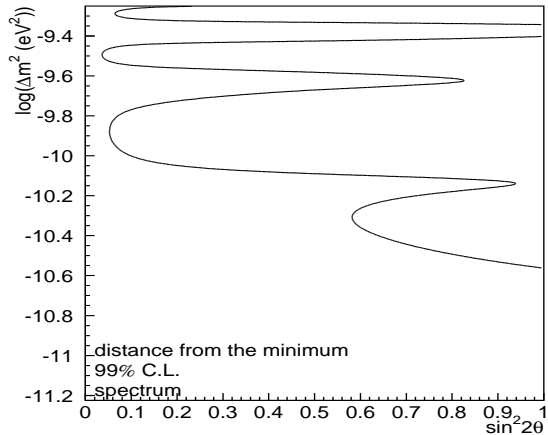}
\end{center}
\vspace{-5mm}
\caption{\em Just-so VO solution: regions allowed by rates (upper panel) and  
excluded by spectrum (low panel)-- courtesy of Y.Totsuka.}
\vspace{-28.1mm}
\end{figure}
\vspace{-40mm}
\newpage

In particular, for $f_B=0.8$ and $f_{Cl}=1.3$ the oscillation 
parameters ($\Delta m^2=4.2 \cdot 10^{-10}$~eV$^2, \sin^22\theta=0.93$)
give a good fit to all rates ($\chi^2$/d.o.f. = 3.0/3) and to the spectrum 
with the excess (see Fig.8).  
More generally, the oscillation parameters give rates in agreement with the
experiments at the $2\sigma$ level when $0.77 \leq f_B \leq 0.83$ and
$1.3 \leq f_{Cl} \leq 1.55$. 

These VO solutions will be referred to as 
HEE VO  (with HEE for High Energy Excess) to distinguish them from ordinary 
just-so VO solutions. 
\vspace{-8mm}
\begin{figure}[h]
\begin{center}
\psfig{bbllx= 100pt, bblly=270pt, bburx=510pt, bbury=660pt,
file=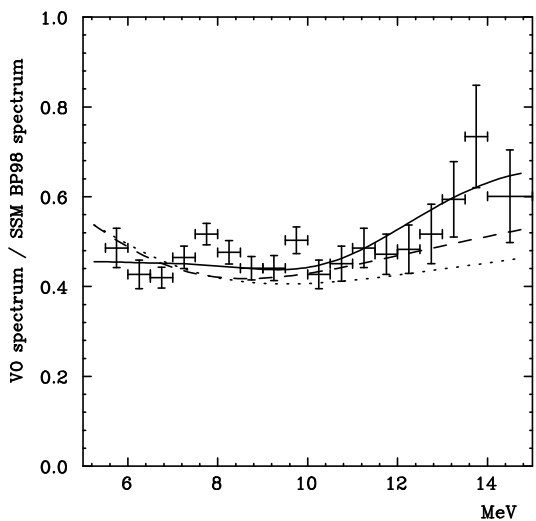, height=7cm , clip=}
\end{center}
\vspace{-13mm}
\caption{\em Ratio of the vacuum oscillation spectra to the SSM spectrum.
The solid curve corresponds to the HEE VO solution with
$\Delta m^2=4.2\cdot 10^{-10}$~eV$^2$ and $\sin^2 2\theta=0.93$.
The dashed and dotted curves correspond to the VO solutions
of Refs.~\cite{HaLa98} and \cite{BKS}, respectively.}
\end{figure}
\vspace{-8mm}
\noindent
The anomalous seasonal variations are rather unusual in the HEE VO solution. 
They are described by time dependence of survival probability for the 
electron neutrino: $P_{\nu_e \to \nu_e}(t)$. 
In particular, for $Be$-neutrinos with energy $E=0.862$~MeV it equals to 
\begin{equation}
P(t)= 1-\sin^2 2\theta
\sin^2\left( \frac{\Delta m^2 a}{4E}\, (1+e\cos \frac{2\pi t}{T} ) \, \right)
\label{seas}
\end{equation}
where $a=1.496\cdot 10^{13}$~cm is the semi-major axis, 
$e=0.01675$ is the eccentricity of the Earth's orbit, and $T=1$~yr is the 
orbital period.

As seen in Fig.9, the case of
the HEE VO (solid curve) is dramatically different from the just-so VO case: 
there are 
two maxima and minima during one year and the survival probability oscillates 
between 
$1-\sin^2 2\theta \approx 0.14$ and 1. The explanation is obvious: the HEE VO 
solution has a large $\Delta m^2$, which results in a phase
$\Delta m^2 a/(4E) \approx 93$, large enough to produce two full harmonics 
during one year, when the phase changes by about 3\% due to the factor
$(1+e\cos 2\pi t/T)$. 

The HEE VO solution predicts (see Fig.9) that $Be$ 
electron neutrinos should arrive almost unsuppressed during about four months 
a year!

According to the SSM, beryllium neutrinos contribute 34.4~SNU out of
the total gallium signal of 129~SNU.
Therefore, the strong $^7$Be neutrino oscillation
predicted by the HEE VO solution also implies an appreciable variation of 
total gallium signal. In Fig.9 the dotted curve shows this variation
for the HEE VO solution, which can be compared with the
weaker variation in the just-so VO solution (dashed-dotted curve).
\vspace{-10mm}
\begin{figure}[hb]
\begin{center}
\psfig{bbllx= 100pt, bblly=270pt, bburx=560pt, bbury=660pt,
file=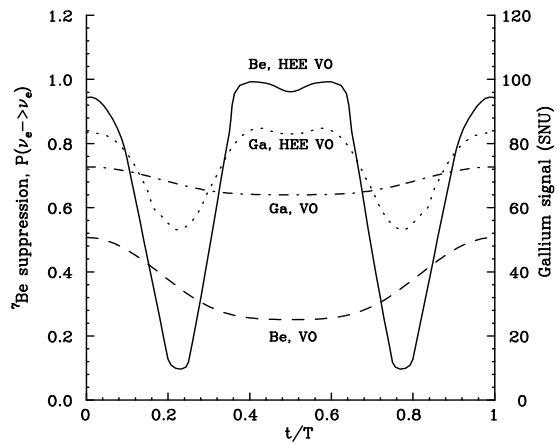, height=6cm,width=7.5cm}
\end{center}
\vspace{-12mm}
\caption{\em Anomalous seasonal variations of the beryllium neutrino flux and 
gallium
signal for the VO and HEE VO solutions.
The survival probability $P_{\nu_e\to \nu_e}$ for Be neutrinos 
is given for the HEE VO (solid curve) and for just-so VO (dashed curve) 
as function of time ($T$ is an orbital period). The dotted
(dash-dotted) curve shows the time variation of gallium signal in SNU for the
HEE VO and for just-so VO~\cite{BKS} solutions.}
\end{figure}
\vspace{-40mm}

\newpage
In Figs. 10 - 15 the predictions of HEE VO solution for seasonal time 
variations are compared 
with observations of GALLEX, SAGE, Homestake and Superkamiokande.
While the agreement of each of the observational data  
with the HEE VO solution might appear accidental and not sta-

\begin{figure}[h]
\begin{center}
\includegraphics[bbllx=105pt,bblly=260pt,bburx=510pt,bbury=660pt,
height=45mm,width=75mm]{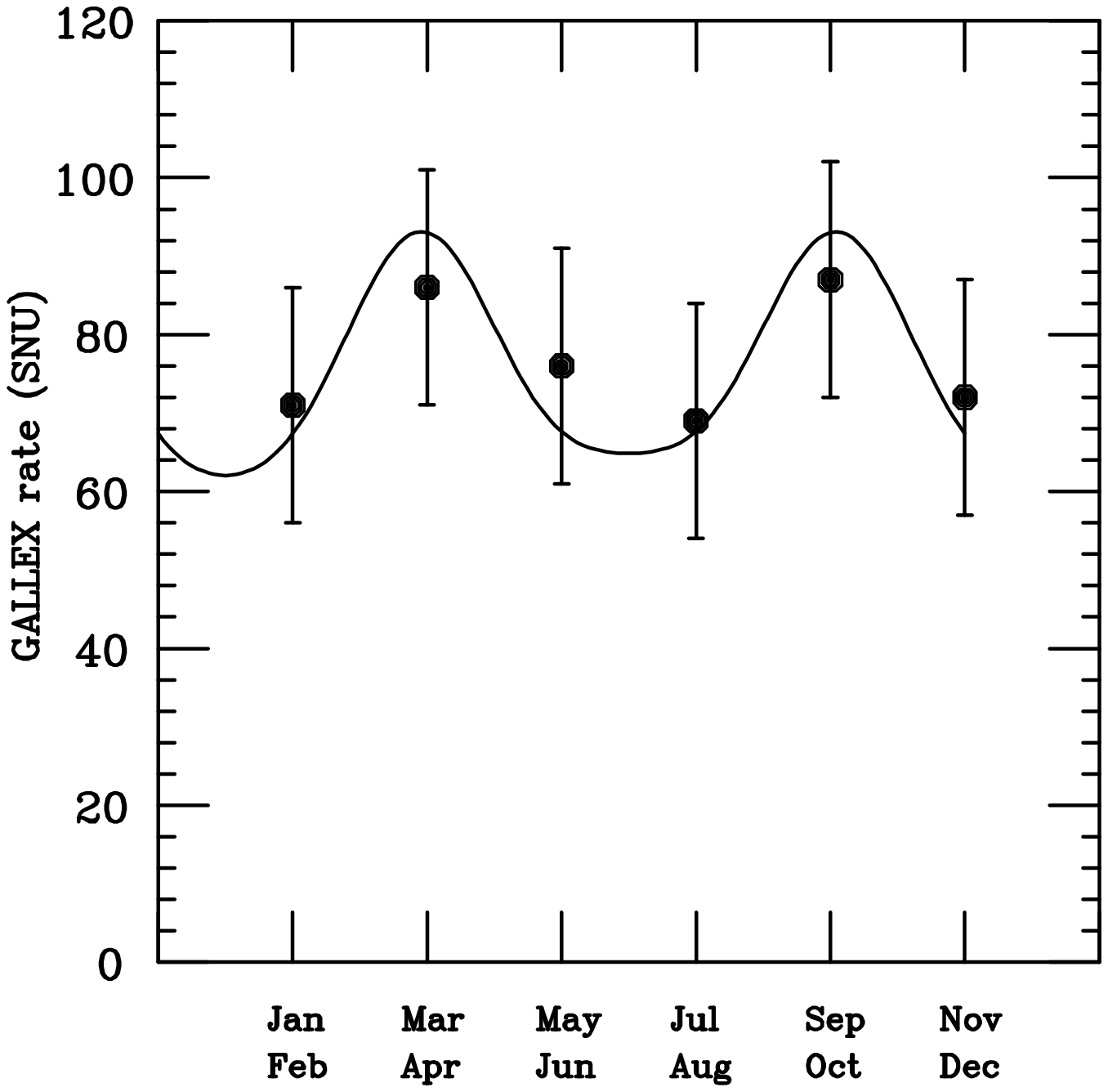}
\end{center}
\vspace{-10mm}
\caption{\em Seasonal variations predicted by the HEE VO in comparison with 
the GALLEX data \cite{GALLEX}. The fit with the HEE VO has 
$\chi^2$/d.o.f.=0.87/4, while a time independent fit gives 
$\chi^2$/d.o.f.=1.36/5.} 
\vspace{-5mm}

\begin{center}
\includegraphics[bbllx=105pt,bblly=260pt,bburx=510pt,bbury=660pt,
height=45mm,width=75mm]{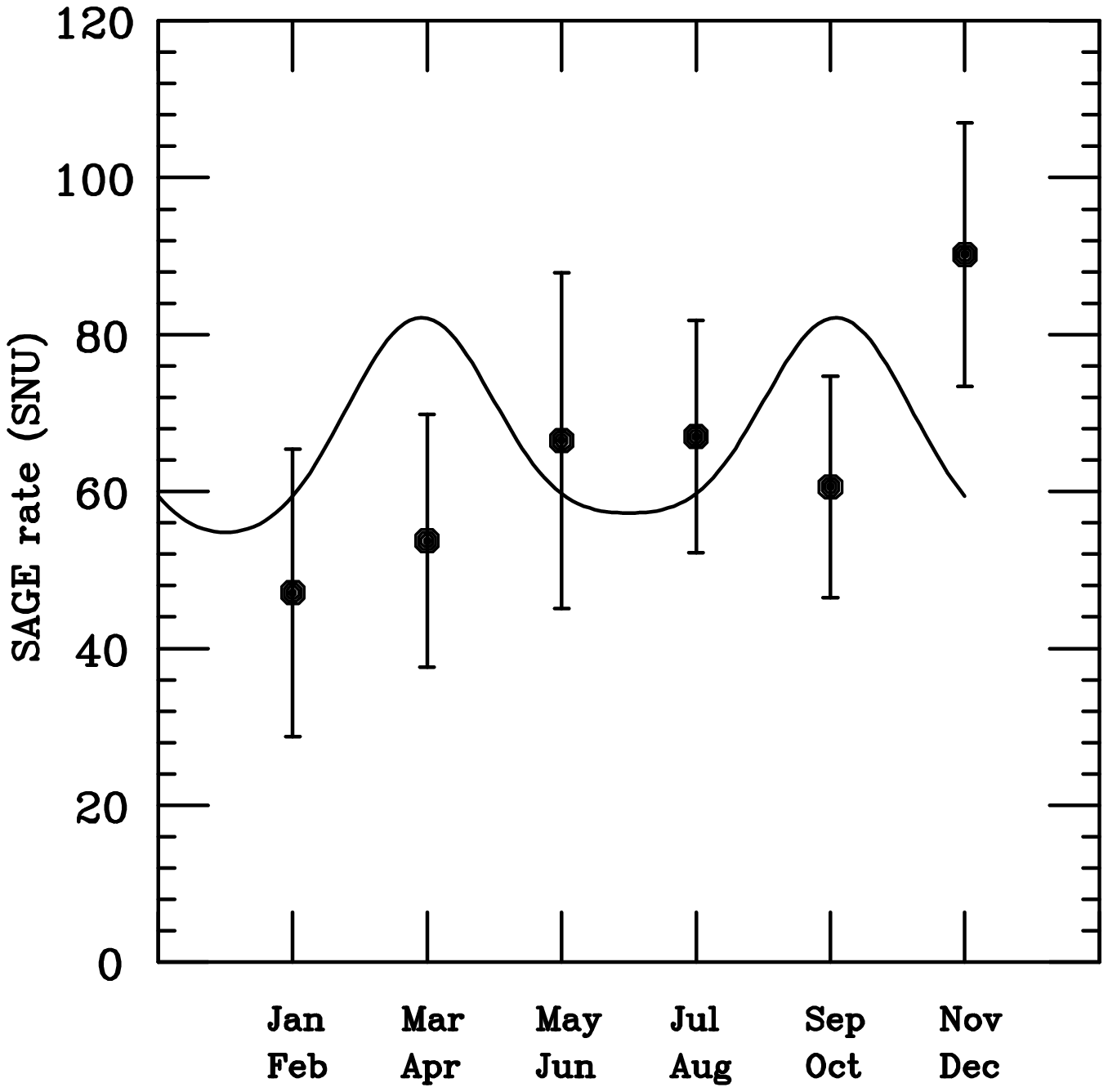}
\end{center}
\vspace{-10mm}
\caption{\em Seasonal variations predicted by the HEE VO in comparison with 
the SAGE preliminary data \cite{SAGE}. The fit with the HEE VO has 
$\chi^2$/d.o.f.=8.9/5, while a time independent fit gives 
$\chi^2$/d.o.f.=3.8/5.}
\end{figure} 
\vspace{-40mm}

\newpage
\noindent
tistically significant, the combined agreement
between this model and experiment   looks like 
indication in favour of the HEE VO solution.
\begin{figure}[ht]
\begin{center}
\includegraphics[bbllx=105pt,bblly=260pt,bburx=510pt,bbury=660pt,
height=43mm,width=75mm]{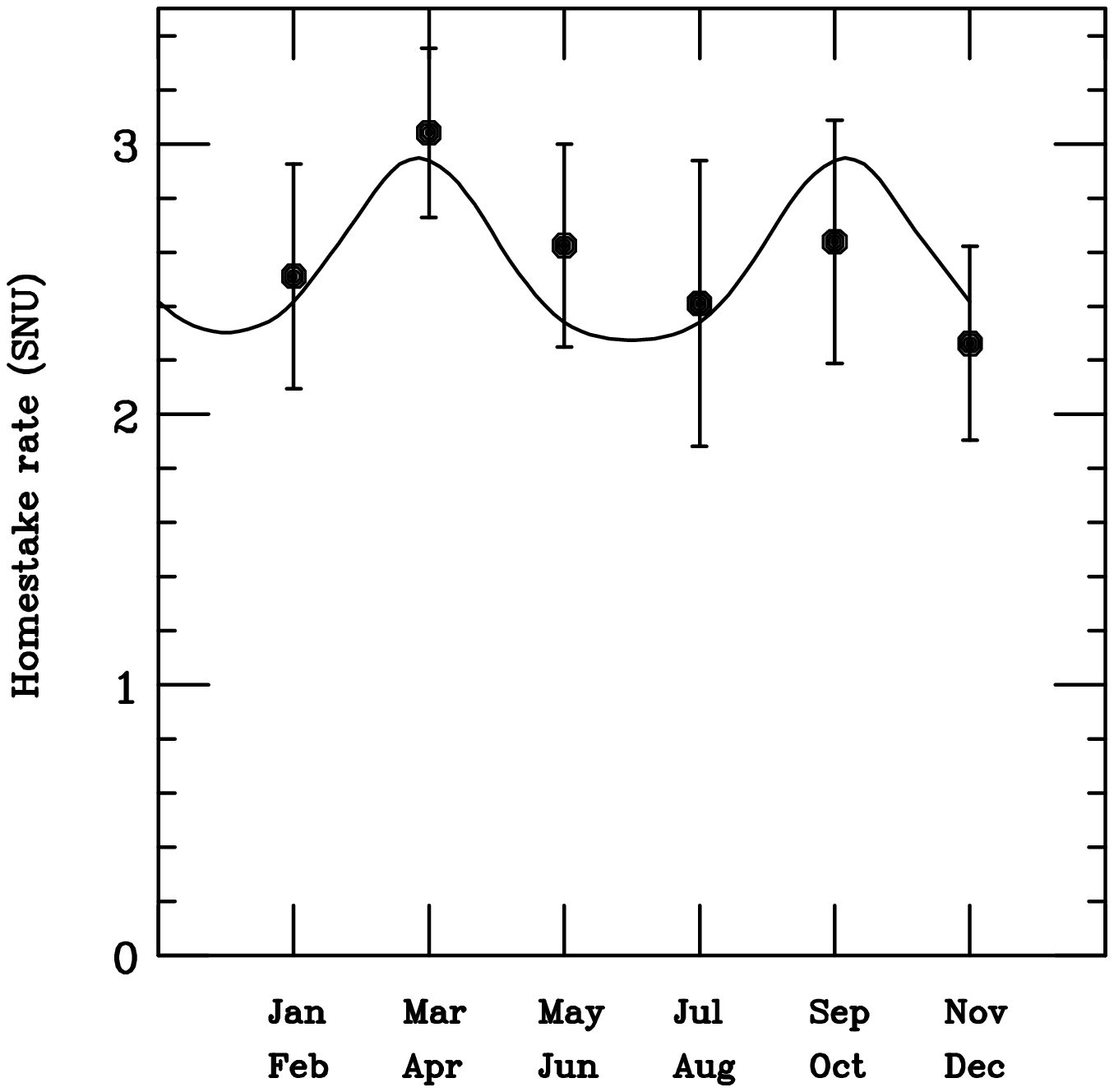}
\end{center}
\vspace{-10mm}
\caption{\em Seasonal variations predicted by the HEE VO in comparison with 
the Homestake data \cite{Hom}. The fit with the HEE VO gives 
$\chi^2$/d.o.f.=1.4/5, while the fit with constant (no-oscillation) gives 
$\chi^2$/d.o.f=3.1/5.}

\begin{center}
\includegraphics[bbllx=100pt,bblly=260pt,bburx=510pt,bbury=654pt,
height=43mm,width=75mm]{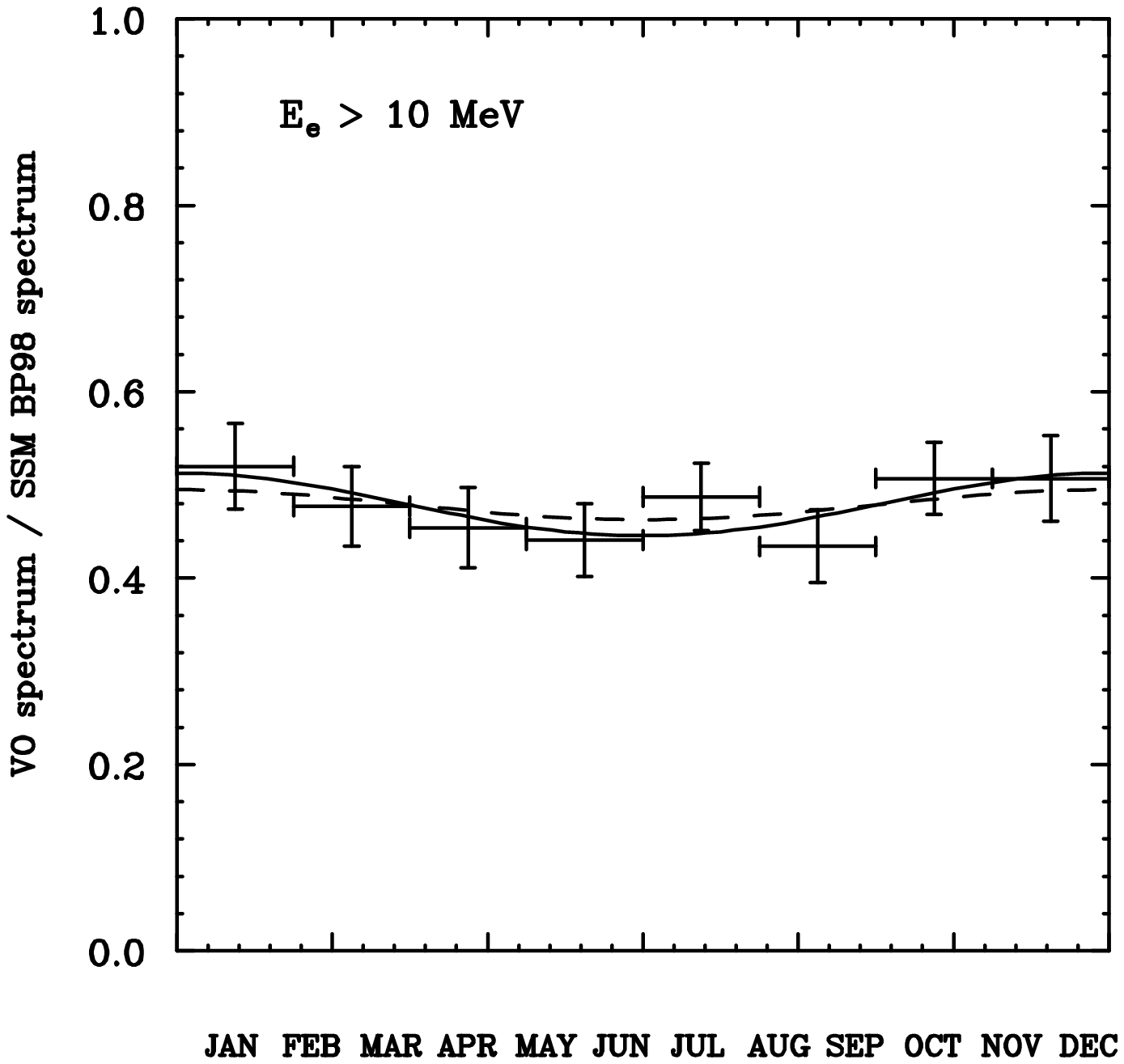}
\end{center}
\vspace{-8mm}
\caption{\em Seasonal variations predicted by the HEE VO for $E_e > 10~MeV$ 
in comparison with the Superkamiokande data at the same energies. The fit with
the HEE VO gives $\chi^2$/d.o.f.=2.7/7, while the one with the geometrical 
effect only gives $\chi^2$/d.o.f.=2.3/7.} 
\end{figure}
\vspace{-40mm}

\newpage
The anomalous seasonal variation of $\nu_{Be}$- flux  predicted by 
the HEE VO will be reliably tested by BOREXINO and LENS.
 
\vspace{-10mm}
\begin{figure}[h]
\begin{center}
\psfig{bbllx= 90pt, bblly=260pt, bburx=530pt, bbury=670pt,
file=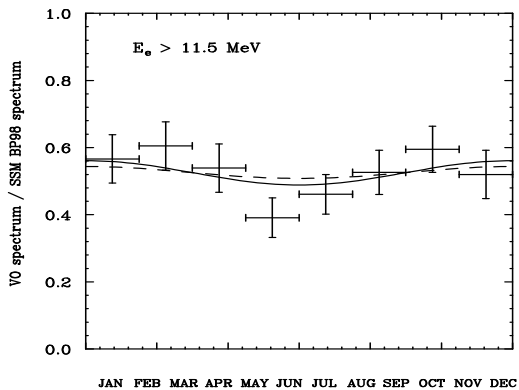, height=5.5cm, width=7.5cm}
\end{center}
\vspace{-10mm}
\caption{\em The same as in Fig.13 for $E_e > 11.5~MeV$. The fit with
the HEE VO gives $\chi^2$/d.o.f.=2.7/7, while the one with the geometrical 
effect only gives $\chi^2$/d.o.f.=2.3/7.}
\vspace{-10mm}
\end{figure}
\vspace{2mm}
{\bf 6.3 RSFP solution}\\*[1mm]
As was recently demonstrated \cite{Nun}, the RSFP solution can successfully 
explain the rates (see also \cite{Babu,Petc}  for early calculations ) and 
high-energy excess in the Superkamiokande spectrum. 

This solution has more free parameters to fit the data. For the Majorana 
neutrino they are: $\Delta m^2$, transition magnetic moment $\mu_{\nu}$, 
scale of 
toroidal magnetic field in the convective zone, $B$, and radial profile for 
magnetic field, $B(r)$, in the wide range of distances. The mixing angle is an 
arbitrary parameter in the RSFP solution which determines the magnetic 
moment, but it must be small enough, $sin 2\theta < 0.25$ \cite{Akh-rev}.

 In Fig.15 the calculated recoil-electron 
spectra (for four magnetic radial profiles) are compared with the 
504d Superkamiokande data. The agreement is reasonably good, though from 
deflections of the individual points one can guess that $\chi^2$ is not 
very small. In Fig.16 the regions explaining
the rates in four solar-neutrino 
experiments are shown in parameter space $\Delta m^2$ and the mean 
magnetic field $<B>$ for two magnetic radial 

\newpage
\noindent
profiles \cite{Nun}.  One can see there the allowed range of parameters.
One of the signatures of the RSFP solution, the time-variation of the 
neutrino signal, is probably testable now. There are two widely discussed 
effects: 11 year periodicity and (June+December)/(March+Sept) ratio of 
fluxes.
\samepage
There are some indications to 11yr periodicity in the Homestake signal, 
especially when correlations with various solar phenomena (sun spot number, 
green coronal line, surface magnetic field {\em etc}) are included. My 
personal opinion is that correlation is allowed as an argument, if the 
the time variability of the signal is established. Such a lesson was taught us 
by a bitter experience with Cyg X-3, when correlation with X-ray variability 
was used as a proof of high-energy gamma-ray signal. Meanwhile, the Homestake 
signal is perfectly compatible \cite{Lis} with constant flux 
($\chi^2/d.o.f. = 0.6$). The data of all other detectors are also consistent 
with time-independent flux.

\vspace{-10mm}  
\begin{figure}[htb]
\epsfxsize=7.5truecm
\centerline{\epsffile{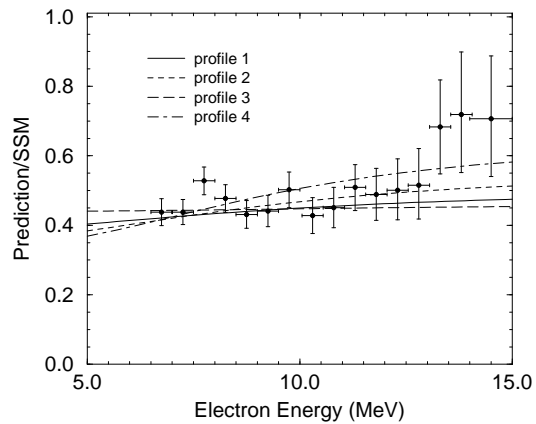}}
\vspace{-10mm}
\caption{\em RSFP spectra for recoil electrons for four different magnetic radial 
profiles, compared with 504d Superkamiokande data.} 
\vspace{-10mm}
\end{figure}

The suppression of neutrino flux in the RSFP model disappears when magnetic 
field vanishes. It happens in two cases: when polarity of magnetic field 
in the Sun changes and when neutrino flux arrives, propagating in the 
plane of solar equator ( June 5 and December 5). This effect is strongest 
for $Be$-neutrinos (see Fig.3).

\newpage
The GALLEX data do not show an excess of the 
rate in June and December or the deficit in March and September. 
Using three month intervals centered at June 5 and December 5 (''high'' rates) 
and at March 5 and September 5 (''low'' rates) the GALLEX collaboration has 
obtained as a mean rate $78.5 \pm 12~SNU$ for 20 ''high'' runs  and 
$90 \pm 12~SNU$ for 19 ''low'' runs, {\em i.e.} within limited statistics 
the wrong-sign effect (T.Kirsten, private communication). In Figs. 13--14 
one can see a similar wrong-sign effect in Superkamiokande data. On the other 
hand there are no accurate model calculations of this effect in the RSFP 
models, to compare with the data above.
\vspace{-15mm}
\begin{figure}[ht]
\epsfxsize=7truecm
\centerline{\epsffile{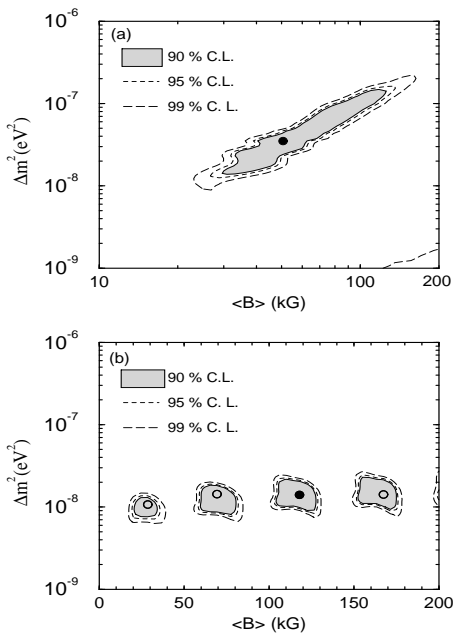}}
\vspace{-15mm}
\caption{\em The RSFP regions allowed by rates for two magnetic field radial 
profiles \cite{Nun}}. 
\end{figure}
\vspace{-13mm}

\section{\bf Future Experiments}

{\em SNO}, e.g. \cite{McD}, is 1 kt heavy water 
detector, which will start to operate this year. In contrast to 
Superkamiokande, electron neutrinos will be
\newpage 
\noindent
directly seen here in the 
CC current reaction $\nu_e +D \to p+p+e^{-}$. Thanks to the large 
cross-section, event rate is close to that of Superkamiokande. 
Neutrino energy is  
given by electron energy and mass difference of $D$ and $H$. The SNO 
data can be helpful in detection of Hep neutrinos above the end of 
$B$-neutrino 
spectrum. The NC reactions $\nu_x+D  \to \nu_x+ p + n$, seen by presence of 
neutrons result in anomalous NC/CC ratio in case of oscillation of 
$\nu_e$ to active neutrino component. This is a signature of neutrino 
oscillation. SNO is more sensitive than Superkamiokande to the day/night 
effect, which is a ``smoking gun'' of MSW effect. This is because $\nu_e$ 
neutrinos are directly measured in CC reaction. Detection of day/night 
effect in Superkamiokande, SNO  and probably ICARUS is the last hope for 
this effect, because all other planned now detectors are not sensitive  
to it.

{\em ICARUS} \cite{ICARUS} is a liquid argon detector. Detection of solar 
neutrinos 
is based on CC-reaction $\nu_e+^{40}\!\!Ar \to ^{40}\!\!K + e^{-}$ and 
$\nu e$ scattering. With excellent energy resolution (about $5\%$) and 
low threshold of electron detection (about $5~MeV$), ICARUS has great potential 
for super-precise measurement of electron spectrum and flux of Hep neutrinos. 

{\em KamLand} \cite{ASuz} is 1kt liquid scintillator detector for 
$\bar{\nu}_e$ neutrinos, based on the Reines reaction 
$\bar{\nu}_e +p \to e^+ + n$. Solar $\bar{\nu}_e$ can be detected, though 
without clear signature of solar origin ({\em e.g.} no directionality). 
Detection of $\bar{\nu}_e$ neutrinos with $E_{\nu} \sim 3~MeV$ from a 
nuclear reactor at distance $L \sim 100~km$ can test the oscillations with 
$\Delta m^2 \sim E_{\nu}/L \sim 3\cdot 10^{-6}~eV^2$, {\em i.e.} close 
to that of LMA MSW solution.  

{\em HELLAZ} \cite{HELLAZ} is a low temperature and high pressure hellium 
detector, registering neutrinos in $\nu e$ scattering. The recoil electron 
energy and scattering angle are measured with high precision and thus 
the energy of neutrino is known with comparable accuracy. This detector is 
designed for $pp$ neutrinos, but probably $Be$ neutrinos can be registered 
too.

{\em BOREXINO} and {LENS} are two low-energy neutrino detectors, 
complementary in physical interpretation of the results. 
BOREXINO will start to operate in the beginning of the next millennium. 
LENS is a new proposal to the Gran Sasso Laboratory based on the recent 
idea put forward by R.S.Raghavan \cite{Ragh}. 

BOREXINO at Gran Sasso is 300t liquid scintillator detector for registering 
$Be$-neutrinos due to $\nu e$ scattering. It measures CC+NC signal 
from $\nu_e$ neutrinos together with NC signal from $\nu_{\mu}$ and 
$\nu_{\tau}$,  
in case of oscillation to  active components. 
LENS is a liquid scintillator detector loaded by Yb or Gd nuclei. Neutrinos 
are detected due to reactions $\nu_e + ^{176}\!Yb \to ^{176}\!Lu^* + e^-$ 
(or $\nu_e + ^{160}\!Gd \to ^{160}\!Tb^* + e^-$) with a threshold 244 keV 
(300 keV). The prompt signal from electron is accompanied by a delayed 
signal from photon or conversion electron from an excited nucleus. This 
strongly reduces the background. LENS will detect $pp$ and 
$Be \; \nu_e$-neutrinos. 
\samepage
The combination of the BOREXINO and LENS data will provide us with the 
following physical information. (i) With $pp$- and $Be$- neutrino fluxes 
measured separately, the whole neutrino spectroscopy of the Sun will be 
completed. The suppression of neutrino fluxes at different energies will 
be explicitly known. (ii) With $Be \; \nu_e$-neutrino flux known from 
LENS, BOREXINO will give a signal from $\nu_{\mu}$ and $\nu_{\tau}$, in 
case of $\nu_e$ oscillation to active neutrinos. Thus, the combination 
of both experiments have a status of {\em appearance oscillation experiment}.
In case of $\nu_e$ oscillation to sterile neutrino, BOREXINO should not 
show the signal in excess of that predicted by LENS. (iii) Just-so VO and 
HEE VO solutions predict strong seasonal variation of $Be$ neutrino   
flux. The BOREXINO/LENS observations will confirm or reject these models.
EIS VO model predicts absence of anomalous seasonal variation accompanied by
suppression 
(by factor $\sim 2$) of both $pp$ and $Be \; \nu_e$ neutrinos. This is also 
can be tested by the combined BOREXINO and LENS data. 
(iv) Both detectors can observe 11yr periodicity in $Be$-neutrino flux and 
measure (June+Dec)/(March+Sept) ratio, which are signatures of the RSFP 
solution.
\vspace{-15mm}
\section{\bf Conclusions}

Solar Neutrino Problem (SNP) is deficit of neutrino fluxes as compared to 
the SSM predictions detected in all solar-neutrino experiments. The 
astrophysical (including nuclear physics) solution to SNP is excluded or 
strongly disfavoured. SNP has a status of disappearance oscillation experiment.
No direct signature of oscillations has been found yet. 

Currently there are six oscillation solutions to SNP: SMA MSW, LMA MSW, 
LOW MSW, Just-so VO, EIS VO, and RSFP. Two of them are disfavoured: 
EIS VO (vacuum oscillations with energy-independent suppression) is excluded 
by observed rates at $99.8 \%$ CL and can survive if the Homestake result is 
excluded from analysis; LOW MSW is seen only at $\geq 99\%$ CL.

Distortion of $B$-neutrino spectrum (as compared with the SSM specrum) is a 
common signature of oscillation solutions (which is absent only in EIS VO 
and weak in the LMA MSW).The ratio of the observed electron spectrum 
(708d of Superkamiokande data) to that of predicted by the SSM model is flat 
at $5.5~MeV \leq E_e \leq 13~MeV$ and has an excess at $E_e \geq 13~MeV$.
This excess cannot be explained by the MSW solutions and by those just-so VO 
solutions, which explain the rates. It is not excluded that this excess is 
due to Hep neutrinos or small systematic experimental error. 

Day/Night effect and zenith-angle dependence of neutrino flux is a signature 
of MSW solutions. After 708d of Superkamiokande observations this effect
(in percent) is $2.9 \pm 1.7 \pm 0.30$, i.e. consistent with zero at 
$1.7 \sigma$.
Statistics, if increased by factor 5, might make this effect statistically 
significant. SNO, in the operation soon, is more sensitive than Superkamiokande 
to day/night effect (due to CC events). There are still chances that 
day/night effect will be discovered in this round of observations. If not, the
future detectors planned at present (BOREXINO, LENS and HELLAZ), will not 
also be able to see it.

The HEE VO solution with $\Delta m^2=4.2\cdot 10^{-10}~eV^2$ and 
$\sin^2 2\theta=0.93$ explains the spectrum with high energy excess and 
the rates, if $B$-neutrino flux is assumed to be $15 - 20 \%$ smaller than in 
SSM and if the chlorine signal is about $30\%$ larger than in the Homestake 
observations. This solution predicts high amplitude semi-annual time 
variation of $Be$-neutrino flux, that can be reliably observed by BOREXINO. 

Another oscillation solution which explains all rates and Superkamiokande 
spectrum (including high energy excess) is the RSFP model. An open problem 
for this model is prediction of 11yr (or 22yr) variations and 
(June+Dec)/(March+Sept) excess, that are not observed.

Future low-energy neutrino detectors, BOREXINO and LENS,  are 
very sensitive to VO solutions and they will either confirm or reject them.
\\[1mm]
{\bf Acknowledgments}\\[1mm]
\samepage
I am grateful to Gianni Fiorentini and Marcello Lissia for enjoyable 
permanent collaboration and discussions. I am very much indebted to 
Yoji Totsuka and Kunio Inoue who provided me with the Superkamiokande data 
in the form of ps-files. I have learned much about Superkamiokande data from 
discussions with Kunio Inoue. I would like to thank Plamen Krastev for 
preparing a compilation of figures from Ref.\cite{BKS} and for discussions. 
Sandro Bettini and Till Kirsten are thanked for discussions and useful 
remarks. 
\samepage
I am honoured and grateful to the organizers of 19th Texas Symposium for 
inviting me for a plenary talk. I appreciate very much their efforts to the 
excellent organization of the conference. 

\samepage


\begin{thebibliography}{99}

\bibitem{Inoue}
K.Inoue, Talk at 8th Int. Workshop ``Neutrino Telescopes'', Venice, 23 - 28 
February, 1999.

\bibitem{GALLEX}
T.Kirsten, Proc. of the 18th Int. Conf. ``Neutrino 98'', 4 - 9 June 1998, to 
be published in Nucl. Phys. B (Proc. Suppl) 1999. 

\bibitem{SAGE}
V.Gavrin,  Proc. of the 18th Int. Conf. ``Neutrino 98'', 4 - 9 June 1998, to 
be published in Nucl. Phys. B (Proc. Suppl) 1999. 
\newpage
\bibitem{Hom}
Homestake collaboration, B~.T.~Cleveland et al, Ap.J., {\bf 496} (1998) 505.

\bibitem{BP98}
J.N.Bahcall, S.Basu and M.H.Pinsonneault, Phys. Lett. {\bf B 433} (1998) 1.

\bibitem{BaBe}
J.N.Bahcall and H.A.Bethe,  Phys. Rev. Lett. {\bf 65}, (1990) 2233.

\bibitem{Blud}
S.Bludman, N.Hata, D.Kennedy and P.Langacker, Phys. Rev. 
{\bf D 47} (1993) 2220.

\bibitem{Cast}
V.Castellani, S.Degl'Innocenti, G.Fiorentini, 
Astron. Astrophys. {\bf 271} (1993) 601.

\bibitem{HaBlLa}
N.Hata, S.Bludman  and P.Langacker, Phys.Rev.{\bf D 49} (1994) 3622.

\bibitem{Be94}
V.Berezinsky, Comm. Nucl. Part. Phys., {\bf 21} (1994) 249.

\bibitem{Ba94}
J.Bahcall, Phys. Lett., {\bf B 338} (1994) 276.

\bibitem{Kw}
W.Kwong and S.P.Rosen, Phys. Rev. Lett. {\bf 73} (1994) 369.

\bibitem{DFL}
S.Degl'Innocenti, G.Fiorentini and M.Lissia, {\bf 43} (1995) 66.

\bibitem{rev}
V.Castellani, S.Degl'Innocenti, G.Fiorentini, M.Lissia and B.Ricci,
Phys. Rep. {\bf 281} (1997) 309.

\bibitem{HaLa98}
N.Hata and P.Langacker, Phys. Rev. {\bf D56} (1997) 6107.

\bibitem{lasthope} 
V.Berezinsky, G.Fiorentini and M.Lissia, Phys. lett. {\bf B365} (1996) 185.

\bibitem{Dz}
W.A.Dziembowski, Bull. Astron. Soc. India, {\bf 24} (1996) 133.

\bibitem{ferrara97}
S.Degl'Innocenti, W.A.Dziembowski, G.Fiorentini and B.Ricci B.,
Astrop. Phys. {\bf 7} (1997) 77.

\bibitem{BPBC}
J.N.Bahcall, M.H.Pinsonneault, S.Basu, and Christensen-Dalsgaard J.,
Phys. Rev. Lett.  {\bf 78} (1997) 171.

\bibitem{Tu-Ch}
A.S.Brun, S.Turck-Chieze and P.Morel, astro-ph/9806272.

\bibitem{ferrara1}
V.Castellani, S.Degl'Innocenti, W.A.Dziembowski, G.Fiorentini and B.Ricci,
Nucl. Phys. B (Proc. Suppl) {\bf 70} (1999) 301 

\bibitem{BKS}
J.N.Bahcall, P.I.Krastev and A.Yu.Smirnov, Phys. Rev. {\bf D58} (1998) 096016.

\bibitem{HCSM}
B.Ricci, V.Berezinsky, S.Degl'Innocenti, W.A.Dziembowski, and G.Fiorentini,
Phys. Lett. {\bf B407} (1997) 155.

\bibitem{Gough}
D.Gough, Annals N.Y. Academy Sci., {\bf 647} (1992) 199.
\newpage
\bibitem{Schatz}
E.Schatzman, in: Proc. 4th Int. Solar Neutrino Conf., Heidelberg, 4 - 11 April 
1997 (ed. W.Hampel) p.21 (1997).

\bibitem{CuHa}
A.Cumming and W.C.Haxton, Phys.Rev.Lett. {\bf77} (1996) 4286.

\bibitem{BFL99}
V.Berezinsky, G.Fiorentini and M.Lissia, astro-ph/9902222.

\bibitem{MSW}
S.P.Mikheyev and A.Yu.Smirnov, Nuovo Cim. {\bf 9 C} (1986) 17,\\
S.P.Mikheyev and A.Yu.Smirnov A.Yu., Sov. Phys. Uspekhi {\bf 30} (1986) 759,\\
L. Wolfenstein,  Phys.Rev. {\bf D17} (1978) 2369.

\bibitem{Pont}
B.Pontecorvo, ZhETP {\bf 33} (1957) 549.

\bibitem{BiPe}
S.M.Bilenky and S.T.Petcov, Rev. Mod. Physics, {\bf 59}, (1987) 671.

\bibitem{Perk}
P.F.Harrison, D.H.Perkins, and W.G.Scott, Phys. Lett. {\bf B 349} (1995) 137;
Phys, Lett. {\bf B 396} (1997) 186.
 
\bibitem{FoVo}
R.Foot and R.R.Volkas, Phys.Rev. {\bf D 52} (1995) 6595.

\bibitem{Comf}
G.Comforto, A.Marcionni, F.N.Martelli, F.Vetrano, M.Lanfranchi, and 
G.Torricelli-Ciamponi, Astrop. Phys. {\bf 5} (1996) 147.

\bibitem{Gold}
A.J.Baltz, A.S.Goldhaber, and M.Goldhaber, Phys. Rev. Lett. {\bf 81} (1998) 
5730.
\bibitem{Akh}
E.Akhmedov, Phys. Lett. {\bf B 213} (1998) 64.

\bibitem{Lim}
C.S.Lim  and  W.J.Marciano, Phys. Rev. {\bf D 37} (1988) 1368.

\bibitem{Akh-rev}
E.Kh.Akhmedov, in: Proc. 4th Int. Solar Neutrino Conf., Heidelberg, 
(ed. W.Hampel) p.388 (1997).
 
\bibitem{VVO}
M.B.Voloshin, M.I.Vysotsky, and L.B.Okun,  Sov. J. Nucl. Phys.
{\bf 44} (1986) 845.

\bibitem{BaFi}
R.Barbieri  and G.Fiorentini, Nucl. Phys. {\bf B 304} (1998) 909.

\bibitem{SchVa}
J.Schechter  and J.W.F.Valle, Phys. Rev. {\bf D 24} (1981) 1883.

\bibitem{Nun}
M.M.Guzzo and H.Nunokawa, hep-ph/9810408.

\bibitem{Valle}
P.C. de Holanda, C.Pena-Garay, M.C.Gonzales-Garcia and J.W.F.Valle,
hep-ph/9903473;\\
J.N.Bahcall, P.I.Krastev and A.Yu.Smirnov, hep-ph/9905220.

\bibitem{Pom}
I.Ya.Pomeranchuk (cited in \cite{BiPe}).

\bibitem{KrPe}
P.I.Krastev and S.T.Petcov, Nucl. Phys. {\bf B449} (1995) 605.

\bibitem{GlKe}
S.L.Glashow, P.J.Kerman, and L.M.Kraus, Phys. Lett. {\bf B190} (1998) 412.
\newpage
\bibitem{MiSm}
S.P.Mikheyev and A.Yu.Smirnov, Phys. Lett. {\bf B 429} (1998) 343.

\bibitem{GeRo}
J.M.Gelb and S.P.Rosen, hep-ph/9809508 (1998).

\bibitem{Sm}
A.Yu.Smirnov, Nucl. Phys. B (Proc. Suppl.) {\bf 70} (1999) 324.

\bibitem{Fo}
B.Faid, G.L.Fogli, E.Lisi and D.Montanino. Astrop. Phys. {\bf 10}
(1999) 93.

\bibitem{MaPe}
M.Maris and S.T.Petcov, hep-ph/9903303.

\bibitem{ASuz}
A.Suzuki, Talk at ''Neutrino 98'', Takayama,4 - 9 June 1998, to be published in
Nucl. Phys. B (Proc. Suppl.) 1999.

\bibitem{Tots}
Y.Totsuka, Talk at 19th Texas Symposium on Relativistic Astrophysics, 
Paris, 14 - 18 December, 1998.

\bibitem{BaKr}
J.N.Bahcall and P.I.Krastev, Phys.Lett. {\bf B 436} (1998) 243,\\
R.Escribano, J.M.Frere, A.Gevaert and D.Monderen, Phys.Lett. {\bf B
444} (1998) 397.

\bibitem{BDFR}
G.Fiorentini, V.Berezinsky, S. Degl'Innocenti, and B.Ricci, Phys. Lett. 
{\bf B 444} (1998) 387.

\bibitem{Lisi}
G.L.Fogli, E.Lisi and D.Montanino, Phys.Lett. {\bf B 434} (1998) 333.

\bibitem{Suz}
Y.Suzuki, Talk at ''Neutrino 98'', Takayama,4 - 9 June 1998, to be published in
Nucl. Phys. B (Proc. Suppl.) 1999.

\bibitem{BFL98}
V.Berezinsky, G.Fiorentini, and M.Lissia, hep-ph/9811352,\\
V.Barger and K.Whisnant, hep-ph/9903262,\\
V.Berezinsky, G.Fiorentitni, and M.Lissia, hep-ph/9904225.

\bibitem{Babu}
K.S.Babu, R.N.Mohapatra and I.Z.Rothstein, Phys. Rev. {\bf D 44} (1991) 2265.

\bibitem{Petc}
E.Kh. Akhmedov, A.Lanza, and S.T.Petcov, Phys. Lett. {\bf B 303} (1993) 85.

\bibitem{Lis}
M.Lissia, Proc. of 18th Texas Symposium on Relativistic Astrophysics, Chicago 
15 - 20 December 1996, World Scientific, eds A.V.Olinto, J.A.Frieman and 
D.N.Schramm, (1998) 706.

\bibitem{McD}
A.B.MacDonald, Nucl. Phys. B (Proc. Suppl.) {\bf 35} (1994) 340.

\bibitem{ICARUS}
ICARUS collaboration, Proposal v.1 and v.2, Laboratori Nazionali del 
Gran Sasso , 1993

\bibitem{HELLAZ}
T.Ypsilantis, Proc. 4th Int. Workshop ''Neutrino Telescopes'',
ed M. Baldo-Ceolin (1992) 289.

\bibitem{BOREXINO}
BOREXINO collaboration, Proposal, Univ. of Milan, 1991.


\bibitem{Ragh}
R.S.Raghavan, Phys. Rev. Lett. {\bf 78} (1997) 3618.
\end{thebibliography}
\end{document}